\documentclass[aps,prl,twocolumn,groupedaddress,amsmath,amssymb,aps]{revtex4-2}
\pdfoutput=1
\usepackage{graphicx}
\usepackage{dcolumn}
\usepackage{bm}
\usepackage{amssymb} 
\usepackage{color,soul}
\usepackage{times,mathptmx}
\usepackage{amsmath}
\usepackage[normalem]{ulem}
\usepackage{soul}

\usepackage{verbatim}

\begin{document}

\title{Microfluidic osmotic compression of a charge-stabilized colloidal dispersion:\\ Equation of state and collective diffusion coefficient}
\author{Camille Keita$^1$}
\author{Yannick Hallez$^2$}
\author{Jean-Baptiste Salmon$^1$}
\affiliation{$^1$CNRS, Solvay, LOF, UMR 5258, Univ. Bordeaux, F-33600 Pessac, France}
\affiliation{$^2$Laboratoire de G\'enie Chimique, Universit\'e de Toulouse, CNRS, INPT, UPS, Toulouse, France}

\date{\today}

\begin{abstract}
We show, using a model coupling mass transport and liquid theory calculations for a charge-stabilized colloidal dispersion, that diffusion significantly limits measurement times  of its Equation Of State (EOS), osmotic pressure vs composition, using the osmotic compression technique.   Following this result, we present a microfluidic chip allowing one to measure the entire EOS of a charged dispersion at the nanoliter scale in a few hours. We also show that time-resolved analyses of relaxation to equilibrium in this microfluidic experiment lead to direct estimates of the collective diffusion coefficient of the dispersion in Donnan equilibrium with a salt reservoir.
\end{abstract}
\maketitle

The dynamics of many out-of-equilibrium processes involving colloidal dispersions, such as drying, ultrafiltration, or sedimentation rely primarily on two fundamental inputs: the Equation of State (EOS)  
 osmotic pressure~$\Pi$  vs colloid concentration~$\varphi$, and the collective diffusion coefficient $D(\varphi)$ relating  the concentration gradient to  the colloid flux in Fick's law~\cite{Russel,Routh:13,Peppin:06,Bowen2007,roa2016ultrafiltration,Piazza1993,Petsev1993}. In ultrafiltration for example, the colloid accumulation at the membrane, the so-called concentration polarization,  limits the process  by the local increase in osmotic pressure, but is limited by  collective diffusion that opposes the formation of concentration gradients~\cite{Belfort:94}. These quantities are also central to many fundamental questions, such as  the validity of a generalized Stokes-Einstein relation for colloidal dispersions~\cite{segre1995viscosity,banchio1999rheology,gupta2015validity}.
 Beyond  dispersions, these quantities, and more particularly the EOS $\Pi$ vs $\varphi$, are also fundamental to probe and understand interactions of soft matter systems at colloidal scales (see e.g., Refs.~\cite{Carrier2007,Yasuda20,Scotti2021}).

\begin{figure}
    \centering
    \includegraphics{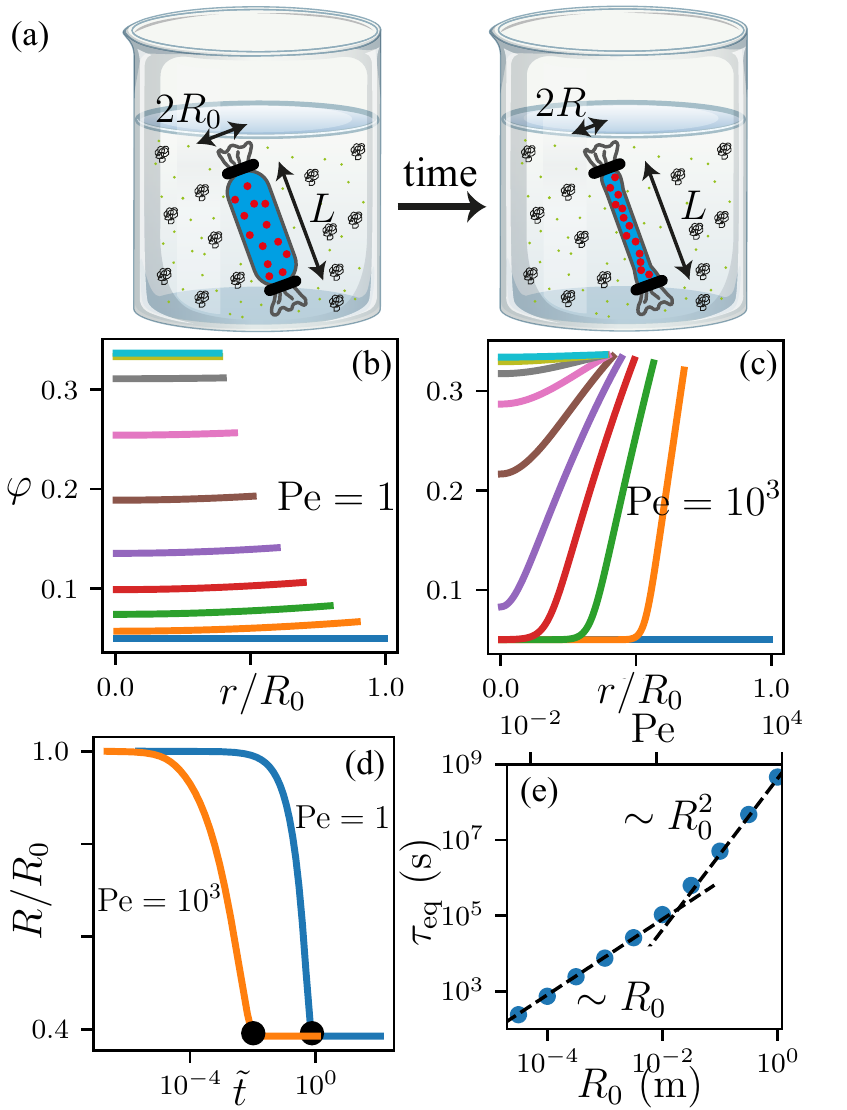}
    \caption{(a)  Osmotic compression for a charged dispersion. (b) and (c)
   Concentration  profiles evenly spaced from $\tilde{t}=0$ to $0.9$ for
      $\text{Pe} = 1$ and from $\tilde{t}=0$ to $0.014$ for
      $\text{Pe} =10^3$.
(d) Corresponding $R/R_0$ vs $\tilde{t}$; $\tilde{\tau}_\text{eq}$ values are shown with bullets.
(e) (\textcolor{blue}{$\bullet$}) $\tau_\text{eq}$  vs $R_0$ estimated from the numerical model. The dashed lines are the scaling laws with $R_0$ as explained in the text. \label{fig:f1}}
\end{figure}

 For charge-stabilized  dispersions, $\Pi(\varphi)$ and  $D(\varphi)$ depend on 
 colloidal interactions~\cite{Belloni2000} and thus on the ion concentration, and
 they are difficult  to measure.  The EOS can be  estimated by various techniques such as analytical ultracentrifugation,
  vapor pressure or membrane osmometry~\cite{Page2008}, but 
 osmotic  compression remains the primary technique because it allows one to measure the EOS at fixed and imposed chemical potentials of salts in order to control colloidal interactions. In this experiment 
 [see Fig.~\ref{fig:f1}(a)],
a dialysis bag  (typical volume $\sim 10~$mL)  filled with a  dispersion is
  immersed in a bath containing salts and polymers that
  impose an osmotic pressure, $\Pi_i$. The bag is not permeable to polymers and colloids but allows the exchange of solvent and ions to balance their chemical potentials, the so-called 
 Donnan  equilibrium. The colloid concentration $\varphi^\star$  in the bag is then measured (often by dry extract)  leading to one point of the EOS, $\Pi_i$ vs $\varphi^\star$, for the conditions imposed by the reservoir. For most dispersions, the long equilibration times ($\geq$ days) prevent  rapid measurements of the EOS by osmotic compression and therefore its use in particular in research and development to quickly screen formulations. Moreover, this technique uses relatively large sample volumes, a few milliliters per measurement point, and is therefore unsuitable for the measurement of EOS of samples that are either expensive or available in too small quantities. 

Measurements of  $D(\varphi)$  are even more challenging for charged dispersions. $D(\varphi)$ is often estimated from dynamic scattering experiments whose clear interpretation is limited to monodisperse colloids diluted enough to avoid multiple scattering~\cite{pusey1978intensity,ackerson1978correlations,pusey1982langevin,Petsev1992,nagele1996dynamics,Gapinski2007}. For less monodisperse and more concentrated colloids, recent works reported estimates of $D(\varphi)$ using  direct measurements of both  concentration gradient and colloid flux in microfluidic  experiments~\cite{Goehring2017,Sobac2020}. Such experiments, however, do not allow one to control the ionic concentration as for the Donnan equilibrium, so that these measurements remain tricky to extrapolate to other processes.    In this work, we first use a model coupling mass transport and liquid theory  to highlight the limiting factor that makes osmotic compression so slow for charged dispersions.
Following this result, we then develop a microfluidic chip  to measure simultaneously  the EOS $\Pi(\varphi)$ of a charge-stabilized dispersion in a few hours and its collective diffusion coefficient $D(\varphi)$, both in Donnan equilibrium.
 
Consider first a cylindrical dialysis bag of  radius $R(t)$ and fixed length $L \gg R$,  filled by an aqueous dispersion of spherical colloids of radius $a$ carrying   $Z$ unit charges [see Fig.~\ref{fig:f1}(a)]. The concentration $c_\text{res}$ of monovalent salt in the bath  is assumed to be fixed.
The shrinkage rate of the bag is given in this  axisymmetrical  model by~\cite{Marbach2019}:
\begin{equation}
    \dot R(t) = \mathcal{L}_p \{\Pi[ \varphi(r=R,t)]-\Pi_i\},
    \label{eq:Rdot}
\end{equation}
where $\mathcal{L}_p$ is the permeability of the membrane. As $R$ decreases over time, both colloid concentration and  osmotic pressure  increase and  equilibrium is reached for a homogeneous concentration $\varphi^\star$, defined by $\Pi( \varphi^\star)=\Pi_i$.  Computing the time $\tau_\text{eq}$ to reach equilibrium  requires solving mass transport  within the bag. For this, we first assume that ions are always at equilibrium because of their high  diffusivity. For local thermodynamic equilibrium conditions (ignoring, e.g., phase nucleation kinetics), assuming isothermal conditions at room temperature $T$ and neglecting advection, the colloid volume fraction  follows:
\begin{equation}
\frac{\partial \varphi}{\partial t} = \frac{1}{r}\frac{\partial}{\partial r}\left(r D(\varphi) \frac{\partial \varphi}{\partial r} \right).
\label{eq:transport}
\end{equation} 
For spherical and monodisperse colloids, $D(\varphi)$ is given by~\cite{banchio2008short}: 
\begin{equation}
    D(\varphi) = D_0 K(\varphi) \chi_{\text{osm}}^{-1},
    \label{eq:defD}
\end{equation}
where $D_0 = (k_B T)/(6\pi \eta a)$ is the Stokes-Einstein coefficient ($\eta$ is the solvent viscosity and $k_B$ is the Boltzmann constant), $K( \varphi)$ is the long-time sedimentation hindrance function, and $\chi_{\text{osm}}$ is the osmotic compressibility coefficient defined by:
\begin{equation}
  \chi_{\text{osm}}^{-1} = \left(\frac{\partial \beta \Pi}{\partial n}\right)_{T,\,c_\text{res}}=\frac{1}{S(q\rightarrow0)},
    \label{eq:ChiOsm}
\end{equation}
where $n = \varphi/(4/3 \pi a^3)$, $\beta=1/(k_B T)$, and $S(q)$ is the colloid structure factor. Note that both 
$\chi_{\text{osm}}$ and $K(\varphi)$ are 
evaluated at fixed reservoir conditions $c_\text{res}$. 
The boundary condition:
\begin{equation}
\varphi(R,t) \dot R  + D[\varphi(R,t)]\frac{\partial \varphi}{\partial r}(R,t) = 0, \label{eq:lim}
\end{equation}
ensures the rejection of the colloids by the membrane.
 As shown in the Supplemental  Material (SM)~\cite{SM}, this model
 can be made dimensionless defining
 $\tilde{r} = r/R_0$ and $\tilde{t} = (D_0/R_0^2) t$, with $R_0 = R(t=0)$, and depends on the P\'eclet number $\text{Pe} = R_0 \mathcal{L}_p \Pi_i/D_0$  comparing the rate of bag shrinkage to colloid diffusion. 
 
To  solve this model,  $\Pi(\varphi)$ and  $D(\varphi)$ are evaluated as follows. 
First, an effective interaction potential between colloids is designed as follows: \begin{equation}
\beta u =l_B Z_\text{eff}^2 \frac{e^{-\kappa_\text{eff} r}}{r} \text{~for~}r\ge 2a,~  \beta u = \infty  \text{~for~} r<2a,  
\label{eq:pot}
\end{equation}
where $l_B$ is the Bjerrum length, and $Z_\text{eff}$ and $\kappa_\text{eff}^{-1}$ are respectively an effective charge and an effective screening length obtained by the Extrapolated Point Charge (EPC) renormalization method~\cite{Boon2015}. The Ornstein-Zernike (OZ) equation is then solved for the structure factor $S(q)$ with either the Rescaled Mean-Spherical Approximation  or the HyperNetted-Chain  closure, which yield indistinguishable results for this  system. $\Pi(\varphi)$  follows from the virial theorem~\cite{Boon2015}.  $D(\varphi)$ is estimated using Eq.~(\ref{eq:defD}), assuming the long-time  $K(\varphi)$ can be approximated by its short-time counterpart $K_s(\varphi)$  evaluated using the renormalized density fluctuation expansion  of Beenakker and Mazur ($\delta\gamma$ method)~\cite{beenakker1984diffusion,BEENAKKER198322,BEENAKKER198448,genz1991collective,banchio2006many,banchio2008short,heinen2011short,westermeier2012structure,riest2015short} taking as sole input the structure factor $S(q)$. 
Any difference between $K_s(\varphi)$ and  $K(\varphi)$ would be due to non-pairwise hydrodynamic interactions, negligible at small $\varphi$ ~\cite{ackerson1978correlations}. As a relevant case, we consider $a=10$~nm, $Z=450$,  $T=20^\circ$C, $\eta = 1$~mPas, 
and  $c_\text{res}=10$~mM. The above calculations  give $D(\varphi)/D_0=3$--16 with  $D_0 \simeq 2.15\times10^{-11}$~m$^2$/s, and $\Pi(\varphi) = 0.13$--80~kPa for $\varphi = 0.05$--0.34 (see Figs.~\ref{fig:EOS} and~\ref{fig:Dphi}). 

Equations~(\ref{eq:Rdot}), (\ref{eq:transport}) and (\ref{eq:lim}) have  been  solved numerically for an initial concentration $\varphi_0=0.05$, 
$\Pi_i = 50$~kPa, and various Pe.
For $\text{Pe} \ll \mathcal{O}(1)$, diffusion homogenizes the concentration   during the shrinkage of the bag as illustrated in Fig.~\ref{fig:f1}(b). 
In this regime, $\varphi(r,t) \simeq \varphi_0 R_0^2/R^2$ owing to mass conservation, and Eq.~(\ref{eq:Rdot}) shows that $\tau_\text{eq} \propto R_0/(\mathcal{L}_p\Pi_0)$ 
[see Fig.~\ref{fig:f1}(e)].
For $\text{Pe} \gg \mathcal{O}(1)$,  Fig.~\ref{fig:f1}(c) shows that  the  concentration close to the membrane quickly tends  to $\varphi^\star$ defined by  $\Pi(\varphi^\star) \simeq \Pi_i$. The  dynamics is then limited by the relaxation by  diffusion of the concentration gradient over the bag, and  $\tau_\text{eq} \propto R_0^2/D_0$ (see  Fig.~\ref{fig:f1}(e) and SM~\cite{SM}). 
Figure~\ref{fig:f1}(e) shows that,  for a permeability of $\mathcal{L}_p = 2 \times 10^{-12}$~m/(Pa s) typical of dialysis bags used for osmotic compression, $\tau_\text{eq} \simeq 1$~day for  $R_0 \simeq 1~$cm.
These results also show that increasing  the permeability $\mathcal{L}_p$  does not decrease equilibration times in this regime
because the transport is diffusion-limited, and  this would only reinforce the importance of the concentration gradient ($\text{Pe} \propto \mathcal{L}_p$).
Figure~\ref{fig:f1}(e) indicates, however,  that reducing the bag size to the microfluidic scale ($R_0 \leq 100~\mu$m) should significantly accelerate osmotic compression.

\begin{figure}
\begin{center}
\includegraphics{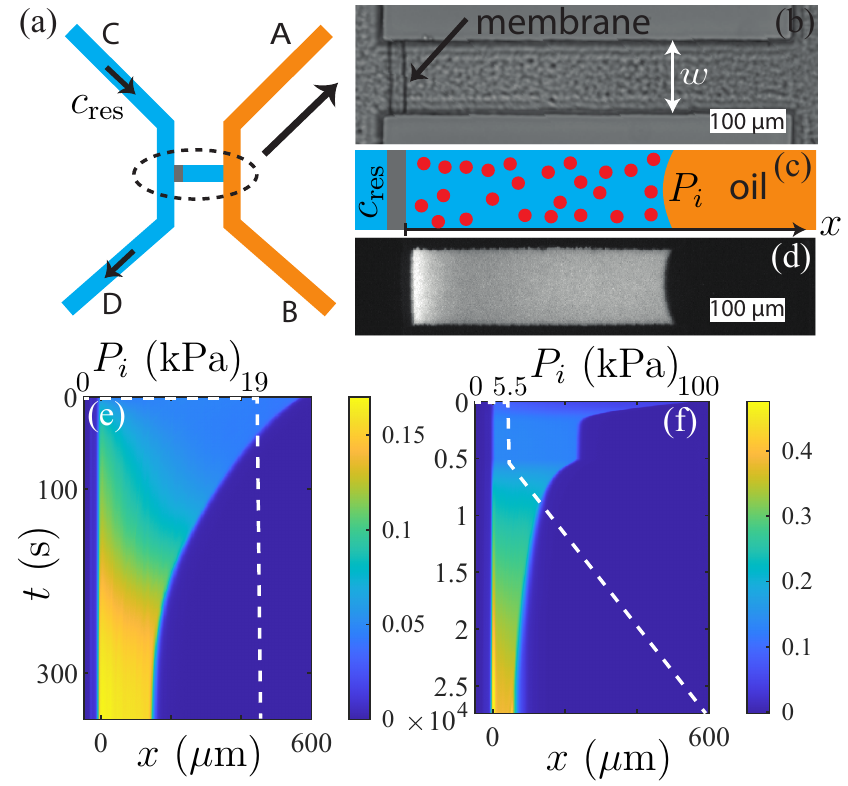}
\caption{(a)  Design of the microfluidic  chip. (b) Bright field view showing the nanoporous membrane in the transverse channel. (c)  Compression of the dispersion by an imposed pressure, $P_i$, and (d) typical fluorescence image. 
(e) Space-time plot of $\varphi(x,t)$  for $P_i = 19~$kPa and $c_\text{res}=1$~mM. (f) Continuous acquisition at 
$c_\text{res}=5$~mM
using a pressure ramp after a constant pressure step. The imposed pressures $P_i$ vs $t$ are shown with white dotted lines on the space-time plots.
\label{fig:setuP}}
\end{center}
\end{figure}

\begin{figure}
\begin{center}
\includegraphics{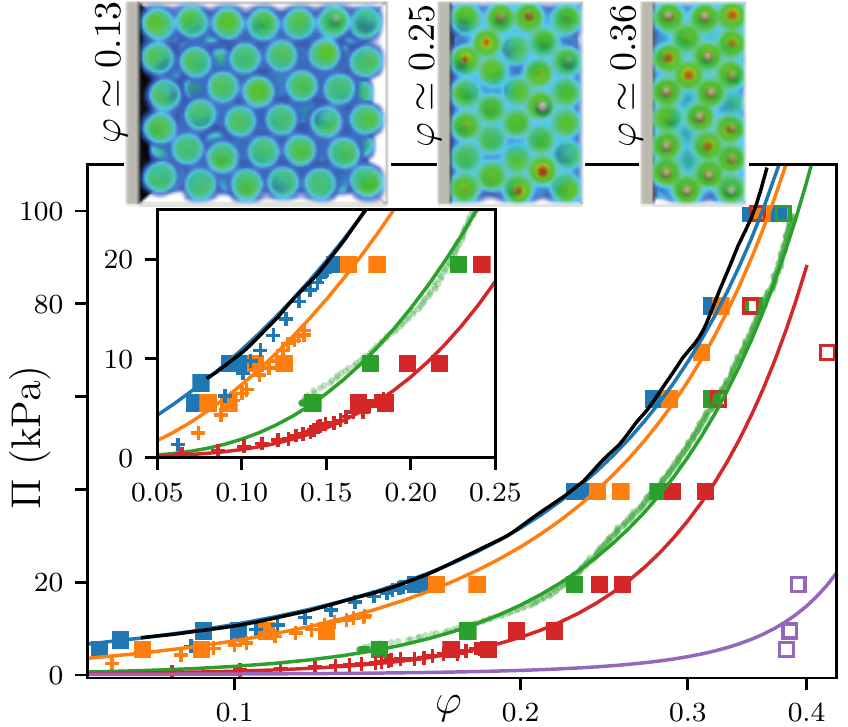}
\caption{Microfluidic acquisition of EOS. Open symbols indicate colloids aggregated irreversibly by the compression. The green dots were obtained from the continuous acquisition shown in Fig.~\ref{fig:setuP}(f). The plus signs correspond to the  out-of-equilibrium measurements using Eq.~(\ref{eq:Ldot}), see text. Salt concentrations are $c_{\text{res}} = 0.1$ (blue), $1$ (orange), $5$ (green), $10$ (red), and $100$ (purple)~mM. The colored lines were obtained from solutions of the OZ equation.  The black line was obtained from PFBD simulations at $c_\text{res}=0.1$ mM. Snapshots are volume renderings of the electric potential field in this simulation at three concentrations. 
\label{fig:EOS}}
\end{center}
\end{figure}

\begin{figure}
\begin{center}
\includegraphics{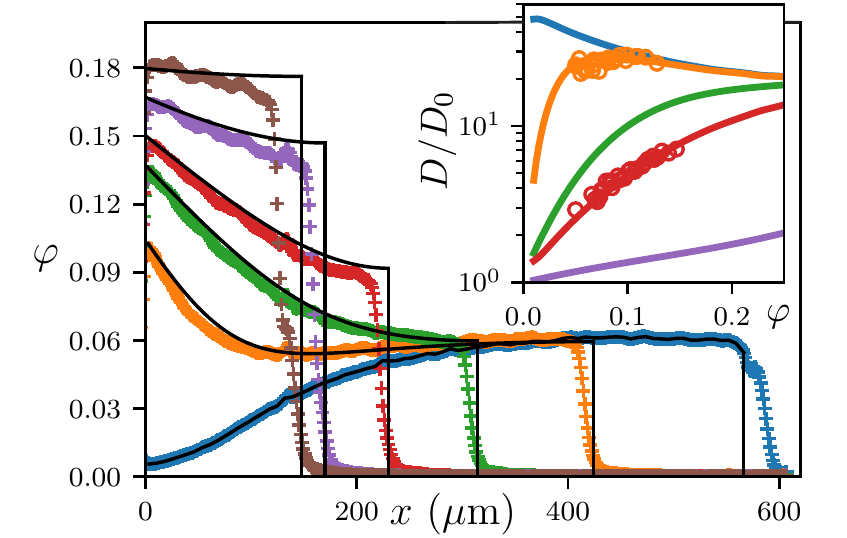}
\caption{ $\varphi(x,t)$ at various $t$ for $P_i = 6~$kPa and $c_\text{res}=10~$mM. Black lines are numerical solutions of  Eqs.~(\ref{eq:Ldot}), (\ref{eq:conv1D}) and (\ref{eq:CL1D})  with $\Pi(\varphi)$ and $D(\varphi)$ calculated from liquid theories. Inset: Theoretical $D(\varphi)/D_0$  at various $c_\text{res}$, same color code as in Fig.~\ref{fig:EOS}. Circles are measurements at $c_\text{res}=1$ and 10~mM. 
\label{fig:Dphi}}
\end{center}
\end{figure}

To measure EOSs in short times, we thus developed a  microfluidic chip mimicking membrane osmometry at the nanoliter scale  to improve mass transport, and  with a reservoir to impose 
a salt chemical potential. First, we made poly(ethylene glycol) diacrylate (PEGDA) chips  using  a rapid prototyping technique~\cite{Rogers2011} (see SM for details~\cite{SM}).  Figure~\ref{fig:setuP}(a) shows their design: two parallel channels (AB and CD) are connected by a transverse channel of height $h = 45~\mu$m, width $w = 100~\mu$m, and length $L = 650~\mu$m.  To integrate a nanoporous membrane, we then inject an aqueous formulation containing PEGDA, a photoinitiator, and poly(ethylene glycol) (PEG) and use spatially-resolved photo-polymerization~\cite{Decock2018} to cross-link a hydrogel at one end of the transverse channel. PEG chains induce a phase separation during polymerization leading to a nanoporous hydrogel~\cite{Lee:2010}. After polymerization, the hydrogel is  flushed leading to the  nanoporous membrane of width $\simeq 25~\mu$m shown in Fig.~\ref{fig:setuP}(b). Such membranes  withstand  pressure drops up to several bars  thanks to the chemical bonds with the acrylate groups of the channel walls, and they
 reject nanoparticles  $\simeq 20$~nm  in diameter while being permeable to water and salts~\cite{Decock2018}. 

For EOS measurements, we studied dispersions of polystyrene particles stabilized by sulfate charges to avoid charge regulation effects
(ThermoFisher, 
S37200, Lot No. 1677598). According to the manufacturer, $2a \simeq~26 \pm 3$~nm and $Z \simeq 450$ from titration. To perform  measurements of concentration profiles, this dispersion is seeded with  a tiny amount (concentration ratio $\simeq 2.7\times 10^{-4}$) of  fluorescent colloids of the same size and the same chemistry (ThermoFisher, 
F8845, Lot No. 2059222). We checked that the fluorescence intensity  varies linearly  with the colloid concentration,  and that photobleaching is negligible thanks to a shutter limiting the overall exposure. 

In a typical experiment,  we  first impose  
$P_C-P_D\simeq 1$~kPa  to ensure a steady  flow ($\simeq 30~\mu$L/h) of an  aqueous sodium chloride (NaCl) solution at concentration $c_\text{res}$ ranging from $0.1$ to 100~mM.
Then, we fill channel AB with the dispersion at a volume fraction of $\simeq 0.075$. Colloids invade the transverse channel by diffusion and  diffusio-phoresis induced by any ion concentration gradient, as  shown in a similar geometry~\cite{Shin2016}.  After enough colloids have invaded the transverse channel,  we flow an inert oil (AR200, Sigma-Aldrich) in channel AB to trap the colloids upstream of the membrane in the volume $h w L \simeq 3$~nL [see Fig.~\ref{fig:setuP}(c)].  
In a typical experiment, we impose a fixed  pressure $P_A = P_B$ at inlets~A and B, and acquire fluorescence images [see Fig.~\ref{fig:setuP}(d) for a typical image]. The pressure drop across the transverse channel, $P_i = P_A-(P_C+P_D)/2$,  ranges from $10$ to 100~kPa, and is  higher than the capillary pressure drop ($\leq 1$~kPa) across the oil/dispersion interface.

Figure~\ref{fig:setuP}(e) shows a space-time plot of the concentration profile $\varphi(x,t)$ estimated from the fluorescence images averaged over the width $w$ for  $P_i = 19~$kPa and $c_\text{res}=1$~mM. The pressure drop compresses the dispersion upstream of the membrane, therefore concentrating the particles. After a transient $\simeq 300$~s, a steady state  is reached with the homogeneous concentration $\varphi^\star \simeq 0.16$.
In such experiments, the thermodynamic equilibrium corresponds to the balance between the imposed mechanical pressure and the osmotic pressure of the dispersion, $\Pi(\varphi^\star) = P_i$, and thus to one point of the EOS, measured here in a few minutes.

To get the entire EOS,  $P_i$ is increased up to 100~kPa by steps of 10~kPa, and  the concentration upstream of the membrane is measured when equilibrium is reached. Figure~\ref{fig:EOS} shows such data for  different reservoirs conditions. The measurement accuracy allows one to highlight  the screening reduction due to a salinity decrease (increase of $\Pi$  at fixed  $\varphi$ for decreasing  $c_\text{res}$), even when approaching the salt-free limit. 
After each point acquisition,  $P_i$ is set to $0$  in order to check the redispersibility of the particles upstream of the membrane (see Fig.~\ref{fig:EOS}). 
The particles  aggregate regardless of the imposed pressure for $c_\text{res} = 100$~mM, but remain redispersible up to $\Pi =100~$kPa for $c_\text{res} \leq 1~$mM.     
This point by point acquisition was also compared to a continuous acquisition in which $P_i$ has been slowly increased up to $P_i = 100$~kPa during $\simeq 6$~h  after an initial step at $P_i =5.5$~kPa. For this experiment at $c_\text{res} = 5$~mM, the slow pressure ramp ensures quasi-equilibrium with almost homogeneous concentration [see Fig.~\ref{fig:setuP}(f)]. As shown in Fig.~\ref{fig:EOS}, the corresponding $\Pi(\varphi)$  curve perfectly overlaps the point-by-point acquisition, demonstrating the possibility to determine  EOS with an unprecedented resolution.  

These experimental EOS have not been compared to measurements made by classical osmotic compression as this would require too large volumes of the commercial dispersion, and would therefore be prohibitively expensive.  Instead, they are compared to the theoretical estimates obtained from the solution of the OZ equation with renormalization in Fig.~\ref{fig:EOS}. Note that this same theoretical framework has already been successfully used to describe EOS measured by classical osmotic compression for a dispersion very close to the system studied here~\cite{Hallez2017}. For the dispersion considered here, perfect agreement over the full range of $ \varphi$ and $c_\text{res}$ is obtained using the measured bare charge  $Z = 450$, and  a unique  colloid radius $a = 10$~nm slightly smaller than the value measured by dynamic light scattering, $a \simeq 11$--$13$~nm with a polydispersity index PdI$=0.1$--0.2 in line with other measurements on the same dispersion, evidencing also a broad size distribution~\cite{Herman2015}.
We believe this small mismatch is primarily due to polydispersity, while calculations strictly  consider monodisperse colloids.

Although  there is no universal model or scaling behavior for the osmotic pressure in the range of volume fraction and salt concentration considered here, a few comments on the physics underlying  the EOS shown in Fig.~\ref{fig:EOS} are in order. The compressibility factor $Z = \Pi/(n k_B T)$ (with $a =10~$nm) reported in Fig.~S2 in the SM~\cite{SM} is always much larger than unity (typically in the range 10--200) because of the significant electrostatic repulsions at the finite volume fractions studied  here. Long-range interactions come into play fairly quickly for all but the highest salt concentration. Therefore, an "effectively concentrated" regime in which the mean intercolloid distance is less than a few times the interaction range is obtained for $\varphi$ greater than a few percents. Under these conditions, colloids are trapped in electrostatic cages, leading to a solidlike (glassy or crystalline) state whose thermodynamic properties are well described by the simple, albeit nonlinear, Poisson-Boltzmann cell model which assumes that the osmotic pressure of the dispersion is dominated by the  contribution of ions~\cite{Alexander1984,Denton2010,Holm2000,trizac_macroion_2007,Hallez2014,Hallez2017}. This model, used with the same physicochemical parameters as those used for the OZ theory, indeed allows one to describe most of the experimental points quite well (see Fig.~S2 in SM~\cite{SM}). However, the high colloidal surface charge and associated significant ion condensation prevent any analytical scaling law from being obtained. Data points at intermediate salt content and volume fraction which are not well described by the cell model correspond to a liquidlike phase in which colloids interact at a  non-vanishing distance (compared to their size). In this regime, the osmotic pressure is dominated by the contribution of direct colloid-colloid interactions and can be estimated from perturbation methods or solutions of the OZ equation (see Ref.~\cite{Hallez2017} for more detailed discussions on the different methods to estimate EOS of charge-stabilized dispersions).

As the use of effective potentials and closures with the OZ equation remains an active research topic especially under concentrated and low salt conditions~\cite{Belloni2000}, we also performed particle scale numerical calculations using Brownian dynamics coupled to a three-dimensional nonlinear Poisson-Boltzmann solver providing the many-body electrostatic forces exerted on each colloid (Particle-Field Brownian Dynamics, PFBD, see SM~\cite{SM})~\cite{fushiki1992molecular,dobnikar2004poisson,Hallez2014,Hallez2016a,Teulon2019}. This method is more precise  as it does not rely on approximate closure relations and does not require the definition of an effective potential. 
As in the  experiments where compression occurs in a single direction, compression is driven along a given axis in the simulation.
Because these numerical calculations are long,  they have been undertaken only for $c_\text{res}=0.1$ mM, the most difficult case to capture with the renormalized OZ approach. The close agreement among simulation, theory, and experiments [see Fig.~\ref{fig:EOS}] confirms again  the robustness of the EPC renormalization used with liquid theories, and also shows that the anisotropy of the compression, as compared to the isotropic case considered with OZ theory, does not play a role in our configuration.

We now exploit the time-resolved concentration profiles during the compression to measure the collective diffusion coefficient $D(\varphi)$.
In the framework of the local thermodynamic equilibrium,
the imposed pressure drop $P_i$ induces a mean flow in the channel given by: 
\begin{eqnarray}
    && J = \mathcal{L}_p\{\Pi[\varphi(x=0,t)]-P_i\}\,, \label{eq:Ldot}
   \end{eqnarray}
because the pressure drop due to the viscous forces  is negligible. In this one-dimensional model, the colloid volume fraction obeys:
\begin{eqnarray}
 && \frac{\partial \varphi}{\partial t} +  J \frac{\partial \varphi}{\partial x} = \frac{\partial}{\partial x}\left(D(\varphi) \frac{\partial \varphi}{\partial x} \right), \label{eq:conv1D}\\
    && J\varphi(x=0,t) = \left(D(\varphi)  \frac{\partial \varphi}{\partial x}\right)_{x=0,\,t},\label{eq:CL1D}
\end{eqnarray}
with Eq.~(\ref{eq:CL1D}) ensuring the  rejection of the colloids by the membrane.
Again in this model, $\Pi(\varphi)$ and $D(\varphi)$ are  assumed in Donnan equilibrium  with the salt concentration $c_\text{res}$ imposed upstream of the membrane. 

We first exploited  $\varphi(x=0,t)$ vs $t$ measured during the first steps of imposed   $P_i$ [see  Fig.~\ref{fig:setuP}(e)], to estimate  the EOS  using Eq.~(\ref{eq:Ldot}) and known values of the permeability $\mathcal{L}_p$. $J \simeq  \dot{A}/w$ is measured from the temporal evolution of the area~$A$ of the dispersion estimated from the fluorescence images [see Fig.~\ref{fig:setuP}(d)]. The data $\Pi[\varphi(x=0,t)]$ vs $\varphi(x=0,t)$ measured this way are shown in  Fig.~\ref{fig:EOS}  with  $\mathcal{L}_p \simeq 1.7 \pm 0.3 \times 10^{-10}$~m/(Pa s) in agreement with values measured using the same technique and only water upstream of the membrane  ($\Pi = 0$). These out-of-equilibrium data  at low $\varphi$ are  in good agreement with the theoretical  EOS at equilibrium thus confirming the hypothesis of local Donnan equilibrium, despite the transmembrane flow.
 
 In a second step, we measured  $\psi(x,t)= \int_0^x \varphi(u,t) \text{d}u$, the amount of colloids trapped upstream of the membrane up to the position $x$. Temporal variations of  $\psi(x,t)$ are related to the  colloid flux at  position $x$ [see Eqs.~(\ref{eq:conv1D}) and (\ref{eq:CL1D})] and can thus be used to estimate $D(\varphi)$ providing we compute $\partial_x \varphi$ and $J$ from the data.
 The inset of Fig.~\ref{fig:Dphi} shows such analysis for various $x$  at $c_\text{res}=1$~mM and 10~mM. 
As expected, $D(\varphi)$  is significantly larger than the Stokes-Einstein $D_0$ due to the electrostatic repulsion, in particular when the dispersion is in equilibrium with a low salinity reservoir. 
These data  are compared  in the inset  of Fig.~\ref{fig:Dphi} to the  theoretical estimates given by the methods explained above, showing a very good agreement.
To further confirm our model, we numerically solved Eqs.~(\ref{eq:Ldot}), (\ref{eq:conv1D}), and 
 (\ref{eq:CL1D}) with $\Pi(\varphi)$ and $D(\varphi)$ estimated theoretically for $c_\text{res}=10$~mM, using the experimental initial condition $\varphi(x,t=0)$ at $P_i=6$~kPa. Figure~\ref{fig:Dphi} shows that the overall dynamics is well predicted by the model, without any free parameters, confirming  the assumption of Donnan equilibrium despite the fluid flow.

In the above comparison, the theoretical $D(\varphi)$ is estimated using the short-time sedimentation hindrance function $K_s(\varphi)$, showing that $K_s \simeq K$ for these charged colloids up to $\varphi \simeq 0.15$. Such a similarity has already been shown numerically for hard-sphere dispersions~\cite{wajnryb2004brownian}, but experimental evidence is still lacking for polydisperse and concentrated dispersions. 
The range of experimental parameters explored above leads to  
unambiguous measurements of  the long-time $D(\varphi)$ only in the  range $\varphi \simeq [0.05$--0.15$]$ and new experiments  are now needed to access a wider $\varphi$ range, especially to investigate  the possible role of many-body hydrodynamics.

In this work, we developed a chip 
to measure with improved  resolution and acquisition time, EOS of charged dispersions in  equilibrium with a salt reservoir down to osmotic pressures of a few kPa.
The dynamics can also be used to measure simultaneously the long-time collective diffusion coefficient of the dispersion, also in Donnan equilibrium, a quantity difficult to measure using classical means. 
In the  future, we plan to use these tools to  address the breakdown of the Donnan equilibrium hypothesis, a subtle issue that could be relevant to many processes, such as ultrafiltration, that involve potentially intense flows.

\begin{acknowledgments}
We thank M. Guirardel, H. Fay, J. Castaing, P. Bacchin and M. Meireles for  fruitful discussions.  This work benefited from meetings within the French working group GDR CNRS 2019 \textit{Solliciter LA Mati\`ere Molle} (SLAMM). We also acknowledge Solvay and the ANR program Grant No. ANR-18-CE06-0021 for financial support. This work was performed using HPC resources from GENCI-CINES/IDRIS (Grant No. A0080911444) and CALMIP (Grant No. 21BP21039R1). 
\end{acknowledgments}


\begin{thebibliography}{54}%
\makeatletter
\providecommand \@ifxundefined [1]{%
 \@ifx{#1\undefined}
}%
\providecommand \@ifnum [1]{%
 \ifnum #1\expandafter \@firstoftwo
 \else \expandafter \@secondoftwo
 \fi
}%
\providecommand \@ifx [1]{%
 \ifx #1\expandafter \@firstoftwo
 \else \expandafter \@secondoftwo
 \fi
}%
\providecommand \natexlab [1]{#1}%
\providecommand \enquote  [1]{``#1''}%
\providecommand \bibnamefont  [1]{#1}%
\providecommand \bibfnamefont [1]{#1}%
\providecommand \citenamefont [1]{#1}%
\providecommand \href@noop [0]{\@secondoftwo}%
\providecommand \href [0]{\begingroup \@sanitize@url \@href}%
\providecommand \@href[1]{\@@startlink{#1}\@@href}%
\providecommand \@@href[1]{\endgroup#1\@@endlink}%
\providecommand \@sanitize@url [0]{\catcode `\\12\catcode `\$12\catcode
  `\&12\catcode `\#12\catcode `\^12\catcode `\_12\catcode `\%12\relax}%
\providecommand \@@startlink[1]{}%
\providecommand \@@endlink[0]{}%
\providecommand \url  [0]{\begingroup\@sanitize@url \@url }%
\providecommand \@url [1]{\endgroup\@href {#1}{\urlprefix }}%
\providecommand \urlprefix  [0]{URL }%
\providecommand \Eprint [0]{\href }%
\providecommand \doibase [0]{https://doi.org/}%
\providecommand \selectlanguage [0]{\@gobble}%
\providecommand \bibinfo  [0]{\@secondoftwo}%
\providecommand \bibfield  [0]{\@secondoftwo}%
\providecommand \translation [1]{[#1]}%
\providecommand \BibitemOpen [0]{}%
\providecommand \bibitemStop [0]{}%
\providecommand \bibitemNoStop [0]{.\EOS\space}%
\providecommand \EOS [0]{\spacefactor3000\relax}%
\providecommand \BibitemShut  [1]{\csname bibitem#1\endcsname}%
\let\auto@bib@innerbib\@empty
\bibitem [{\citenamefont {Russel}\ \emph {et~al.}(1989)\citenamefont {Russel},
  \citenamefont {Saville},\ and\ \citenamefont {Schowalter}}]{Russel}%
  \BibitemOpen
  \bibfield  {author} {\bibinfo {author} {\bibfnamefont {W.~B.}\ \bibnamefont
  {Russel}}, \bibinfo {author} {\bibfnamefont {D.~A.}\ \bibnamefont
  {Saville}},\ and\ \bibinfo {author} {\bibfnamefont {W.~R.}\ \bibnamefont
  {Schowalter}},\ }\href@noop {} {\emph {\bibinfo {title} {Colloidal
  dispersions}}}\ (\bibinfo  {publisher} {Cambridge University Press},\
  \bibinfo {year} {1989})\BibitemShut {NoStop}%
\bibitem [{\citenamefont {Routh}(2013)}]{Routh:13}%
  \BibitemOpen
  \bibfield  {author} {\bibinfo {author} {\bibfnamefont {A.~F.}\ \bibnamefont
  {Routh}},\ }\bibfield  {title} {\bibinfo {title} {Drying of thin colloidal
  films},\ }\href@noop {} {\bibfield  {journal} {\bibinfo  {journal} {Rep.
  Prog. Phys.}\ }\textbf {\bibinfo {volume} {76}},\ \bibinfo {pages} {046603}
  (\bibinfo {year} {2013})}\BibitemShut {NoStop}%
\bibitem [{\citenamefont {Peppin}\ \emph {et~al.}(2006)\citenamefont {Peppin},
  \citenamefont {Elliott},\ and\ \citenamefont {Worster}}]{Peppin:06}%
  \BibitemOpen
  \bibfield  {author} {\bibinfo {author} {\bibfnamefont {S.~S.}\ \bibnamefont
  {Peppin}}, \bibinfo {author} {\bibfnamefont {J.~A.}\ \bibnamefont
  {Elliott}},\ and\ \bibinfo {author} {\bibfnamefont {M.~G.}\ \bibnamefont
  {Worster}},\ }\bibfield  {title} {\bibinfo {title} {Solidification of
  colloidal suspensions},\ }\href@noop {} {\bibfield  {journal} {\bibinfo
  {journal} {J. Fluid Mech.}\ }\textbf {\bibinfo {volume} {554}},\ \bibinfo
  {pages} {147} (\bibinfo {year} {2006})}\BibitemShut {NoStop}%
\bibitem [{\citenamefont {Bowen}\ and\ \citenamefont
  {Williams}(2007)}]{Bowen2007}%
  \BibitemOpen
  \bibfield  {author} {\bibinfo {author} {\bibfnamefont {W.~R.}\ \bibnamefont
  {Bowen}}\ and\ \bibinfo {author} {\bibfnamefont {P.~M.}\ \bibnamefont
  {Williams}},\ }\bibfield  {title} {\bibinfo {title} {{Quantitative predictive
  modelling of ultrafiltration processes: Colloidal science approaches}},\
  }\href@noop {} {\bibfield  {journal} {\bibinfo  {journal} {Adv. Colloid
  Interface Sci.}\ }\textbf {\bibinfo {volume} {134-135}},\ \bibinfo {pages}
  {3} (\bibinfo {year} {2007})}\BibitemShut {NoStop}%
\bibitem [{\citenamefont {Roa}\ \emph {et~al.}(2016)\citenamefont {Roa},
  \citenamefont {Menne}, \citenamefont {Riest}, \citenamefont {Buzatu},
  \citenamefont {Zholkovskiy}, \citenamefont {Dhont}, \citenamefont
  {Wessling},\ and\ \citenamefont {N{\"a}gele}}]{roa2016ultrafiltration}%
  \BibitemOpen
  \bibfield  {author} {\bibinfo {author} {\bibfnamefont {R.}~\bibnamefont
  {Roa}}, \bibinfo {author} {\bibfnamefont {D.}~\bibnamefont {Menne}}, \bibinfo
  {author} {\bibfnamefont {J.}~\bibnamefont {Riest}}, \bibinfo {author}
  {\bibfnamefont {P.}~\bibnamefont {Buzatu}}, \bibinfo {author} {\bibfnamefont
  {E.~K.}\ \bibnamefont {Zholkovskiy}}, \bibinfo {author} {\bibfnamefont
  {J.~K.~G.}\ \bibnamefont {Dhont}}, \bibinfo {author} {\bibfnamefont
  {M.}~\bibnamefont {Wessling}},\ and\ \bibinfo {author} {\bibfnamefont
  {G.}~\bibnamefont {N{\"a}gele}},\ }\bibfield  {title} {\bibinfo {title}
  {Ultrafiltration of charge-stabilized dispersions at low salinity},\
  }\href@noop {} {\bibfield  {journal} {\bibinfo  {journal} {Soft Matter}\
  }\textbf {\bibinfo {volume} {12}},\ \bibinfo {pages} {4638} (\bibinfo {year}
  {2016})}\BibitemShut {NoStop}%
\bibitem [{\citenamefont {Piazza}\ \emph {et~al.}(1993)\citenamefont {Piazza},
  \citenamefont {Bellini},\ and\ \citenamefont {Degiorgio}}]{Piazza1993}%
  \BibitemOpen
  \bibfield  {author} {\bibinfo {author} {\bibfnamefont {R.}~\bibnamefont
  {Piazza}}, \bibinfo {author} {\bibfnamefont {T.}~\bibnamefont {Bellini}},\
  and\ \bibinfo {author} {\bibfnamefont {V.}~\bibnamefont {Degiorgio}},\
  }\bibfield  {title} {\bibinfo {title} {{Equilibrium sedimentation profile of
  screened charged colloids: a test of the Hard-Sphere equation of state}},\
  }\href@noop {} {\bibfield  {journal} {\bibinfo  {journal} {Phys. Rev. Lett.}\
  }\textbf {\bibinfo {volume} {71}},\ \bibinfo {pages} {4267} (\bibinfo {year}
  {1993})}\BibitemShut {NoStop}%
\bibitem [{\citenamefont {Petsev}\ \emph {et~al.}(1993)\citenamefont {Petsev},
  \citenamefont {Starov},\ and\ \citenamefont {Ivanov}}]{Petsev1993}%
  \BibitemOpen
  \bibfield  {author} {\bibinfo {author} {\bibfnamefont {D.~N.}\ \bibnamefont
  {Petsev}}, \bibinfo {author} {\bibfnamefont {V.~M.}\ \bibnamefont {Starov}},\
  and\ \bibinfo {author} {\bibfnamefont {I.~B.}\ \bibnamefont {Ivanov}},\
  }\bibfield  {title} {\bibinfo {title} {{Concentrated dispersions of charged
  colloidal particles: Sedimentation, ultrafiltration and diffusion}},\
  }\href@noop {} {\bibfield  {journal} {\bibinfo  {journal} {Colloids Surf.,
  A}\ }\textbf {\bibinfo {volume} {81}},\ \bibinfo {pages} {65} (\bibinfo
  {year} {1993})}\BibitemShut {NoStop}%
\bibitem [{\citenamefont {Belfort}\ \emph {et~al.}(1994)\citenamefont
  {Belfort}, \citenamefont {Davis},\ and\ \citenamefont {Zydney}}]{Belfort:94}%
  \BibitemOpen
  \bibfield  {author} {\bibinfo {author} {\bibfnamefont {G.}~\bibnamefont
  {Belfort}}, \bibinfo {author} {\bibfnamefont {R.}~\bibnamefont {Davis}},\
  and\ \bibinfo {author} {\bibfnamefont {A.~L.}\ \bibnamefont {Zydney}},\
  }\bibfield  {title} {\bibinfo {title} {The behavior of suspensions and
  macromolecular solutions in crossflow microfiltration},\ }\href@noop {}
  {\bibfield  {journal} {\bibinfo  {journal} {J. Membrane Sci.}\ }\textbf
  {\bibinfo {volume} {96}},\ \bibinfo {pages} {1} (\bibinfo {year}
  {1994})}\BibitemShut {NoStop}%
\bibitem [{\citenamefont {Segre}\ \emph {et~al.}(1995)\citenamefont {Segre},
  \citenamefont {Meeker}, \citenamefont {Pusey},\ and\ \citenamefont
  {Poon}}]{segre1995viscosity}%
  \BibitemOpen
  \bibfield  {author} {\bibinfo {author} {\bibfnamefont {P.~N.}\ \bibnamefont
  {Segre}}, \bibinfo {author} {\bibfnamefont {S.~P.}\ \bibnamefont {Meeker}},
  \bibinfo {author} {\bibfnamefont {P.~N.}\ \bibnamefont {Pusey}},\ and\
  \bibinfo {author} {\bibfnamefont {W.~C.~K.}\ \bibnamefont {Poon}},\
  }\bibfield  {title} {\bibinfo {title} {Viscosity and structural relaxation in
  suspensions of hard-sphere colloids},\ }\href@noop {} {\bibfield  {journal}
  {\bibinfo  {journal} {Phys. Rev. Lett.}\ }\textbf {\bibinfo {volume} {75}},\
  \bibinfo {pages} {958} (\bibinfo {year} {1995})}\BibitemShut {NoStop}%
\bibitem [{\citenamefont {Banchio}\ \emph {et~al.}(1999)\citenamefont
  {Banchio}, \citenamefont {Bergenholtz},\ and\ \citenamefont
  {N{\"a}gele}}]{banchio1999rheology}%
  \BibitemOpen
  \bibfield  {author} {\bibinfo {author} {\bibfnamefont {A.~J.}\ \bibnamefont
  {Banchio}}, \bibinfo {author} {\bibfnamefont {J.}~\bibnamefont
  {Bergenholtz}},\ and\ \bibinfo {author} {\bibfnamefont {G.}~\bibnamefont
  {N{\"a}gele}},\ }\bibfield  {title} {\bibinfo {title} {Rheology and dynamics
  of colloidal suspensions},\ }\href@noop {} {\bibfield  {journal} {\bibinfo
  {journal} {Phys. Rev. Lett.}\ }\textbf {\bibinfo {volume} {82}},\ \bibinfo
  {pages} {1792} (\bibinfo {year} {1999})}\BibitemShut {NoStop}%
\bibitem [{\citenamefont {Gupta}\ \emph {et~al.}(2015)\citenamefont {Gupta},
  \citenamefont {Stellbrink}, \citenamefont {Zaccarelli}, \citenamefont
  {Likos}, \citenamefont {Camargo}, \citenamefont {Holmqvist}, \citenamefont
  {Allgaier}, \citenamefont {Willner},\ and\ \citenamefont
  {Richter}}]{gupta2015validity}%
  \BibitemOpen
  \bibfield  {author} {\bibinfo {author} {\bibfnamefont {S.}~\bibnamefont
  {Gupta}}, \bibinfo {author} {\bibfnamefont {J.}~\bibnamefont {Stellbrink}},
  \bibinfo {author} {\bibfnamefont {E.}~\bibnamefont {Zaccarelli}}, \bibinfo
  {author} {\bibfnamefont {C.~N.}\ \bibnamefont {Likos}}, \bibinfo {author}
  {\bibfnamefont {M.}~\bibnamefont {Camargo}}, \bibinfo {author} {\bibfnamefont
  {P.}~\bibnamefont {Holmqvist}}, \bibinfo {author} {\bibfnamefont
  {J.}~\bibnamefont {Allgaier}}, \bibinfo {author} {\bibfnamefont
  {L.}~\bibnamefont {Willner}},\ and\ \bibinfo {author} {\bibfnamefont
  {D.}~\bibnamefont {Richter}},\ }\bibfield  {title} {\bibinfo {title}
  {Validity of the stokes-einstein relation in soft colloids up to the glass
  transition},\ }\href@noop {} {\bibfield  {journal} {\bibinfo  {journal}
  {Phys. Rev. Lett.}\ }\textbf {\bibinfo {volume} {115}},\ \bibinfo {pages}
  {128302} (\bibinfo {year} {2015})}\BibitemShut {NoStop}%
\bibitem [{\citenamefont {Carri\`ere}\ \emph {et~al.}(2007)\citenamefont
  {Carri\`ere}, \citenamefont {Page}, \citenamefont {Dubois}, \citenamefont
  {Zemb}, \citenamefont {C\"olfen}, \citenamefont {Meister}, \citenamefont
  {Belloni}, \citenamefont {Sch\"onhoff},\ and\ \citenamefont
  {M\"ohwald}}]{Carrier2007}%
  \BibitemOpen
  \bibfield  {author} {\bibinfo {author} {\bibfnamefont {D.}~\bibnamefont
  {Carri\`ere}}, \bibinfo {author} {\bibfnamefont {M.}~\bibnamefont {Page}},
  \bibinfo {author} {\bibfnamefont {M.}~\bibnamefont {Dubois}}, \bibinfo
  {author} {\bibfnamefont {T.}~\bibnamefont {Zemb}}, \bibinfo {author}
  {\bibfnamefont {H.}~\bibnamefont {C\"olfen}}, \bibinfo {author}
  {\bibfnamefont {A.}~\bibnamefont {Meister}}, \bibinfo {author} {\bibfnamefont
  {L.}~\bibnamefont {Belloni}}, \bibinfo {author} {\bibfnamefont
  {M.}~\bibnamefont {Sch\"onhoff}},\ and\ \bibinfo {author} {\bibfnamefont
  {H.}~\bibnamefont {M\"ohwald}},\ }\bibfield  {title} {\bibinfo {title}
  {{Osmotic pressure in colloid science: clay dispersions, catanionics,
  polyelectrolyte complexes and polyelectrolyte multilayers}},\ }\href@noop {}
  {\bibfield  {journal} {\bibinfo  {journal} {Colloids Surf., A}\ }\textbf
  {\bibinfo {volume} {303}},\ \bibinfo {pages} {137} (\bibinfo {year}
  {2007})}\BibitemShut {NoStop}%
\bibitem [{\citenamefont {Yasuda}\ \emph {et~al.}(2020)\citenamefont {Yasuda},
  \citenamefont {Sakumichi}, \citenamefont {Chung},\ and\ \citenamefont
  {Sakai}}]{Yasuda20}%
  \BibitemOpen
  \bibfield  {author} {\bibinfo {author} {\bibfnamefont {T.}~\bibnamefont
  {Yasuda}}, \bibinfo {author} {\bibfnamefont {N.}~\bibnamefont {Sakumichi}},
  \bibinfo {author} {\bibfnamefont {U.-i.}\ \bibnamefont {Chung}},\ and\
  \bibinfo {author} {\bibfnamefont {T.}~\bibnamefont {Sakai}},\ }\bibfield
  {title} {\bibinfo {title} {Universal equation of state describes osmotic
  pressure throughout gelation process},\ }\href@noop {} {\bibfield  {journal}
  {\bibinfo  {journal} {Phys. Rev. Lett.}\ }\textbf {\bibinfo {volume} {125}},\
  \bibinfo {pages} {267801} (\bibinfo {year} {2020})}\BibitemShut {NoStop}%
\bibitem [{\citenamefont {Scotti}\ \emph {et~al.}(2021)\citenamefont {Scotti},
  \citenamefont {Pelaez-Fernandez}, \citenamefont {Gasser},\ and\ \citenamefont
  {Fernandez-Nieves}}]{Scotti2021}%
  \BibitemOpen
  \bibfield  {author} {\bibinfo {author} {\bibfnamefont {A.}~\bibnamefont
  {Scotti}}, \bibinfo {author} {\bibfnamefont {M.}~\bibnamefont
  {Pelaez-Fernandez}}, \bibinfo {author} {\bibfnamefont {U.}~\bibnamefont
  {Gasser}},\ and\ \bibinfo {author} {\bibfnamefont {A.}~\bibnamefont
  {Fernandez-Nieves}},\ }\bibfield  {title} {\bibinfo {title} {{Osmotic
  pressure of suspensions comprised of charged microgels}},\ }\href@noop {}
  {\bibfield  {journal} {\bibinfo  {journal} {Phys. Rev. E}\ }\textbf {\bibinfo
  {volume} {103}},\ \bibinfo {pages} {012609} (\bibinfo {year}
  {2021})}\BibitemShut {NoStop}%
\bibitem [{\citenamefont {Belloni}(2000)}]{Belloni2000}%
  \BibitemOpen
  \bibfield  {author} {\bibinfo {author} {\bibfnamefont {L.}~\bibnamefont
  {Belloni}},\ }\bibfield  {title} {\bibinfo {title} {{Colloidal
  interactions}},\ }\href@noop {} {\bibfield  {journal} {\bibinfo  {journal}
  {J. Phys.: Condens. Matter}\ }\textbf {\bibinfo {volume} {12}},\ \bibinfo
  {pages} {R549} (\bibinfo {year} {2000})}\BibitemShut {NoStop}%
\bibitem [{\citenamefont {Page}\ \emph {et~al.}(2008)\citenamefont {Page},
  \citenamefont {Zemb}, \citenamefont {Dubois},\ and\ \citenamefont
  {C{\"{o}}lfen}}]{Page2008}%
  \BibitemOpen
  \bibfield  {author} {\bibinfo {author} {\bibfnamefont {M.~G.}\ \bibnamefont
  {Page}}, \bibinfo {author} {\bibfnamefont {T.}~\bibnamefont {Zemb}}, \bibinfo
  {author} {\bibfnamefont {M.}~\bibnamefont {Dubois}},\ and\ \bibinfo {author}
  {\bibfnamefont {H.}~\bibnamefont {C{\"{o}}lfen}},\ }\bibfield  {title}
  {\bibinfo {title} {{Osmotic Pressure and Phase Boundary Determination of
  Multiphase Systems by Analytical Ultracentrifugation}},\ }\href@noop {}
  {\bibfield  {journal} {\bibinfo  {journal} {ChemPhysChem}\ }\textbf {\bibinfo
  {volume} {9}},\ \bibinfo {pages} {882} (\bibinfo {year} {2008})}\BibitemShut
  {NoStop}%
\bibitem [{\citenamefont {Pusey}(1978)}]{pusey1978intensity}%
  \BibitemOpen
  \bibfield  {author} {\bibinfo {author} {\bibfnamefont {P.~N.}\ \bibnamefont
  {Pusey}},\ }\bibfield  {title} {\bibinfo {title} {Intensity fluctuation
  spectroscopy of charged brownian particles: the coherent scattering
  function},\ }\href@noop {} {\bibfield  {journal} {\bibinfo  {journal} {J.
  Phys. A: Math. Gen.}\ }\textbf {\bibinfo {volume} {11}},\ \bibinfo {pages}
  {119} (\bibinfo {year} {1978})}\BibitemShut {NoStop}%
\bibitem [{\citenamefont {Ackerson}(1978)}]{ackerson1978correlations}%
  \BibitemOpen
  \bibfield  {author} {\bibinfo {author} {\bibfnamefont {B.~J.}\ \bibnamefont
  {Ackerson}},\ }\bibfield  {title} {\bibinfo {title} {Correlations for
  interacting brownian particles. {II}},\ }\href@noop {} {\bibfield  {journal}
  {\bibinfo  {journal} {J. Chem. Phys.}\ }\textbf {\bibinfo {volume} {69}},\
  \bibinfo {pages} {684} (\bibinfo {year} {1978})}\BibitemShut {NoStop}%
\bibitem [{\citenamefont {Pusey}\ and\ \citenamefont
  {Tough}(1982)}]{pusey1982langevin}%
  \BibitemOpen
  \bibfield  {author} {\bibinfo {author} {\bibfnamefont {P.~N.}\ \bibnamefont
  {Pusey}}\ and\ \bibinfo {author} {\bibfnamefont {R.~J.~A.}\ \bibnamefont
  {Tough}},\ }\bibfield  {title} {\bibinfo {title} {Langevin approach to the
  dynamics of interacting brownian particles},\ }\href@noop {} {\bibfield
  {journal} {\bibinfo  {journal} {J. Phys. A: Math. Gen.}\ }\textbf {\bibinfo
  {volume} {15}},\ \bibinfo {pages} {1291} (\bibinfo {year}
  {1982})}\BibitemShut {NoStop}%
\bibitem [{\citenamefont {Petsev}\ and\ \citenamefont
  {Denkov}(1992)}]{Petsev1992}%
  \BibitemOpen
  \bibfield  {author} {\bibinfo {author} {\bibfnamefont {D.~N.}\ \bibnamefont
  {Petsev}}\ and\ \bibinfo {author} {\bibfnamefont {N.~D.}\ \bibnamefont
  {Denkov}},\ }\bibfield  {title} {\bibinfo {title} {{Diffusion of charged
  colloidal particles at low volume fraction: Theoretical model and light
  scattering experiments}},\ }\href@noop {} {\bibfield  {journal} {\bibinfo
  {journal} {J. Colloid Interface Sci.}\ }\textbf {\bibinfo {volume} {149}},\
  \bibinfo {pages} {329} (\bibinfo {year} {1992})}\BibitemShut {NoStop}%
\bibitem [{\citenamefont {N{\"a}gele}(1996)}]{nagele1996dynamics}%
  \BibitemOpen
  \bibfield  {author} {\bibinfo {author} {\bibfnamefont {G.}~\bibnamefont
  {N{\"a}gele}},\ }\bibfield  {title} {\bibinfo {title} {On the dynamics and
  structure of charge-stabilized suspensions},\ }\href@noop {} {\bibfield
  {journal} {\bibinfo  {journal} {Phys. Rep.}\ }\textbf {\bibinfo {volume}
  {272}},\ \bibinfo {pages} {215} (\bibinfo {year} {1996})}\BibitemShut
  {NoStop}%
\bibitem [{\citenamefont {Gapinski}\ \emph {et~al.}(2007)\citenamefont
  {Gapinski}, \citenamefont {Patkowski}, \citenamefont {Banchio}, \citenamefont
  {Holmqvist}, \citenamefont {Meier}, \citenamefont {Lettinga},\ and\
  \citenamefont {N\"agele}}]{Gapinski2007}%
  \BibitemOpen
  \bibfield  {author} {\bibinfo {author} {\bibfnamefont {J.}~\bibnamefont
  {Gapinski}}, \bibinfo {author} {\bibfnamefont {A.}~\bibnamefont {Patkowski}},
  \bibinfo {author} {\bibfnamefont {A.~J.}\ \bibnamefont {Banchio}}, \bibinfo
  {author} {\bibfnamefont {P.}~\bibnamefont {Holmqvist}}, \bibinfo {author}
  {\bibfnamefont {G.}~\bibnamefont {Meier}}, \bibinfo {author} {\bibfnamefont
  {M.~P.}\ \bibnamefont {Lettinga}},\ and\ \bibinfo {author} {\bibfnamefont
  {G.}~\bibnamefont {N\"agele}},\ }\bibfield  {title} {\bibinfo {title}
  {{Collective diffusion in charge-stabilized suspensions: concentration and
  salt effects}},\ }\href@noop {} {\bibfield  {journal} {\bibinfo  {journal}
  {J. Chem. Phys.}\ }\textbf {\bibinfo {volume} {126}},\ \bibinfo {pages}
  {104905} (\bibinfo {year} {2007})}\BibitemShut {NoStop}%
\bibitem [{\citenamefont {Goehring}\ \emph {et~al.}(2017)\citenamefont
  {Goehring}, \citenamefont {Li},\ and\ \citenamefont
  {Kiatkirakajorn}}]{Goehring2017}%
  \BibitemOpen
  \bibfield  {author} {\bibinfo {author} {\bibfnamefont {L.}~\bibnamefont
  {Goehring}}, \bibinfo {author} {\bibfnamefont {J.}~\bibnamefont {Li}},\ and\
  \bibinfo {author} {\bibfnamefont {P.-C.}\ \bibnamefont {Kiatkirakajorn}},\
  }\bibfield  {title} {\bibinfo {title} {{Drying paint: from micro-scale
  dynamics to mechanical instabilities}},\ }\href@noop {} {\bibfield  {journal}
  {\bibinfo  {journal} {Philos. Trans. R. Soc., A}\ }\textbf {\bibinfo {volume}
  {375}},\ \bibinfo {pages} {2093} (\bibinfo {year} {2017})}\BibitemShut
  {NoStop}%
\bibitem [{\citenamefont {Sobac}\ \emph {et~al.}(2020)\citenamefont {Sobac},
  \citenamefont {Dehaeck}, \citenamefont {Bouchaudy},\ and\ \citenamefont
  {Salmon}}]{Sobac2020}%
  \BibitemOpen
  \bibfield  {author} {\bibinfo {author} {\bibfnamefont {B.}~\bibnamefont
  {Sobac}}, \bibinfo {author} {\bibfnamefont {S.}~\bibnamefont {Dehaeck}},
  \bibinfo {author} {\bibfnamefont {A.}~\bibnamefont {Bouchaudy}},\ and\
  \bibinfo {author} {\bibfnamefont {J.-B.}\ \bibnamefont {Salmon}},\ }\bibfield
   {title} {\bibinfo {title} {{Collective diffusion coefficient of a charged
  colloidal dispersion: Interferometric measurements in a drying drop}},\
  }\href@noop {} {\bibfield  {journal} {\bibinfo  {journal} {Soft Matter}\
  }\textbf {\bibinfo {volume} {16}},\ \bibinfo {pages} {8213} (\bibinfo {year}
  {2020})}\BibitemShut {NoStop}%
\bibitem [{\citenamefont {Marbach}\ and\ \citenamefont
  {Bocquet}(2019)}]{Marbach2019}%
  \BibitemOpen
  \bibfield  {author} {\bibinfo {author} {\bibfnamefont {S.}~\bibnamefont
  {Marbach}}\ and\ \bibinfo {author} {\bibfnamefont {L.}~\bibnamefont
  {Bocquet}},\ }\bibfield  {title} {\bibinfo {title} {{Osmosis, from molecular
  insights to large-scale applications}},\ }\href@noop {} {\bibfield  {journal}
  {\bibinfo  {journal} {Chem. Soc. Rev.}\ }\textbf {\bibinfo {volume} {48}},\
  \bibinfo {pages} {3102} (\bibinfo {year} {2019})}\BibitemShut {NoStop}%
\bibitem [{\citenamefont {Banchio}\ and\ \citenamefont
  {N{\"a}gele}(2008)}]{banchio2008short}%
  \BibitemOpen
  \bibfield  {author} {\bibinfo {author} {\bibfnamefont {A.~J.}\ \bibnamefont
  {Banchio}}\ and\ \bibinfo {author} {\bibfnamefont {G.}~\bibnamefont
  {N{\"a}gele}},\ }\bibfield  {title} {\bibinfo {title} {Short-time transport
  properties in dense suspensions: from neutral to charge-stabilized colloidal
  spheres},\ }\href@noop {} {\bibfield  {journal} {\bibinfo  {journal} {J.
  Chem. Phys.}\ }\textbf {\bibinfo {volume} {128}},\ \bibinfo {pages} {104903}
  (\bibinfo {year} {2008})}\BibitemShut {NoStop}%
\bibitem [{SM()}]{SM}%
  \BibitemOpen
  \bibinfo {note} {See Supplemental Material at
  http://link.aps.org/supplemental/ 10.1103/PhysRevE.104.L062601 for the
  dimensionless transport model; more details on liquid theory calculations,
  PEGDA chip prototyping, and PFBD simulations; and more discussion on the
  compressibility factor. The Supplemental Material also includes
  Refs.~\cite{Salmon2017humidity,liu2000boundary}}\BibitemShut {NoStop}%
\bibitem [{\citenamefont {Boon}\ \emph {et~al.}(2015)\citenamefont {Boon},
  \citenamefont {Guerrero-Garc{\'\i}a}, \citenamefont {{van Roij}},\ and\
  \citenamefont {{de la Cruz}}}]{Boon2015}%
  \BibitemOpen
  \bibfield  {author} {\bibinfo {author} {\bibfnamefont {N.}~\bibnamefont
  {Boon}}, \bibinfo {author} {\bibfnamefont {G.~I.}\ \bibnamefont
  {Guerrero-Garc{\'\i}a}}, \bibinfo {author} {\bibfnamefont {R.}~\bibnamefont
  {{van Roij}}},\ and\ \bibinfo {author} {\bibfnamefont {M.~O.}\ \bibnamefont
  {{de la Cruz}}},\ }\bibfield  {title} {\bibinfo {title} {Effective charges
  and virial pressure of concentrated macroion solutions},\ }\href@noop {}
  {\bibfield  {journal} {\bibinfo  {journal} {Proc. Natl. Acad. Sci. USA}\
  }\textbf {\bibinfo {volume} {112}},\ \bibinfo {pages} {9242} (\bibinfo {year}
  {2015})}\BibitemShut {NoStop}%
\bibitem [{\citenamefont {Beenakker}\ and\ \citenamefont
  {Mazur}(1984)}]{beenakker1984diffusion}%
  \BibitemOpen
  \bibfield  {author} {\bibinfo {author} {\bibfnamefont {C.~W.~J.}\
  \bibnamefont {Beenakker}}\ and\ \bibinfo {author} {\bibfnamefont
  {P.}~\bibnamefont {Mazur}},\ }\bibfield  {title} {\bibinfo {title} {Diffusion
  of spheres in a concentrated suspension {II}},\ }\href@noop {} {\bibfield
  {journal} {\bibinfo  {journal} {Physica A}\ }\textbf {\bibinfo {volume}
  {126}},\ \bibinfo {pages} {349} (\bibinfo {year} {1984})}\BibitemShut
  {NoStop}%
\bibitem [{\citenamefont {Beenakker}\ and\ \citenamefont
  {Mazur}(1983)}]{BEENAKKER198322}%
  \BibitemOpen
  \bibfield  {author} {\bibinfo {author} {\bibfnamefont {C.~W.~J.}\
  \bibnamefont {Beenakker}}\ and\ \bibinfo {author} {\bibfnamefont
  {P.}~\bibnamefont {Mazur}},\ }\bibfield  {title} {\bibinfo {title} {Diffusion
  of spheres in a concentrated suspension: Resummation of many-body
  hydrodynamic interactions},\ }\href@noop {} {\bibfield  {journal} {\bibinfo
  {journal} {Phys. Lett. A}\ }\textbf {\bibinfo {volume} {98}},\ \bibinfo
  {pages} {22} (\bibinfo {year} {1983})}\BibitemShut {NoStop}%
\bibitem [{\citenamefont {Beenakker}(1984)}]{BEENAKKER198448}%
  \BibitemOpen
  \bibfield  {author} {\bibinfo {author} {\bibfnamefont {C.~W.~J.}\
  \bibnamefont {Beenakker}},\ }\bibfield  {title} {\bibinfo {title} {The
  effective viscosity of a concentrated suspension of spheres (and its relation
  to diffusion)},\ }\href@noop {} {\bibfield  {journal} {\bibinfo  {journal}
  {Physica A}\ }\textbf {\bibinfo {volume} {128}},\ \bibinfo {pages} {48}
  (\bibinfo {year} {1984})}\BibitemShut {NoStop}%
\bibitem [{\citenamefont {Genz}\ and\ \citenamefont
  {Klein}(1991)}]{genz1991collective}%
  \BibitemOpen
  \bibfield  {author} {\bibinfo {author} {\bibfnamefont {U.}~\bibnamefont
  {Genz}}\ and\ \bibinfo {author} {\bibfnamefont {R.}~\bibnamefont {Klein}},\
  }\bibfield  {title} {\bibinfo {title} {Collective diffusion of charged
  spheres in the presence of hydrodynamic interaction},\ }\href@noop {}
  {\bibfield  {journal} {\bibinfo  {journal} {Physica A}\ }\textbf {\bibinfo
  {volume} {171}},\ \bibinfo {pages} {26} (\bibinfo {year} {1991})}\BibitemShut
  {NoStop}%
\bibitem [{\citenamefont {Banchio}\ \emph {et~al.}(2006)\citenamefont
  {Banchio}, \citenamefont {Gapinski}, \citenamefont {Patkowski}, \citenamefont
  {H\"au\ss{}ler}, \citenamefont {Fluerasu}, \citenamefont {Sacanna},
  \citenamefont {Holmqvist}, \citenamefont {Meier}, \citenamefont {Lettinga},\
  and\ \citenamefont {N\"agele}}]{banchio2006many}%
  \BibitemOpen
  \bibfield  {author} {\bibinfo {author} {\bibfnamefont {A.~J.}\ \bibnamefont
  {Banchio}}, \bibinfo {author} {\bibfnamefont {J.}~\bibnamefont {Gapinski}},
  \bibinfo {author} {\bibfnamefont {A.}~\bibnamefont {Patkowski}}, \bibinfo
  {author} {\bibfnamefont {W.}~\bibnamefont {H\"au\ss{}ler}}, \bibinfo {author}
  {\bibfnamefont {A.}~\bibnamefont {Fluerasu}}, \bibinfo {author}
  {\bibfnamefont {S.}~\bibnamefont {Sacanna}}, \bibinfo {author} {\bibfnamefont
  {P.}~\bibnamefont {Holmqvist}}, \bibinfo {author} {\bibfnamefont
  {G.}~\bibnamefont {Meier}}, \bibinfo {author} {\bibfnamefont {M.~P.}\
  \bibnamefont {Lettinga}},\ and\ \bibinfo {author} {\bibfnamefont
  {G.}~\bibnamefont {N\"agele}},\ }\bibfield  {title} {\bibinfo {title}
  {Many-body hydrodynamic interactions in charge-stabilized suspensions},\
  }\href@noop {} {\bibfield  {journal} {\bibinfo  {journal} {Phys. Rev. Lett.}\
  }\textbf {\bibinfo {volume} {96}},\ \bibinfo {pages} {138303} (\bibinfo
  {year} {2006})}\BibitemShut {NoStop}%
\bibitem [{\citenamefont {Heinen}\ \emph {et~al.}(2011)\citenamefont {Heinen},
  \citenamefont {Banchio},\ and\ \citenamefont {N{\"a}gele}}]{heinen2011short}%
  \BibitemOpen
  \bibfield  {author} {\bibinfo {author} {\bibfnamefont {M.}~\bibnamefont
  {Heinen}}, \bibinfo {author} {\bibfnamefont {A.~J.}\ \bibnamefont
  {Banchio}},\ and\ \bibinfo {author} {\bibfnamefont {G.}~\bibnamefont
  {N{\"a}gele}},\ }\bibfield  {title} {\bibinfo {title} {Short-time rheology
  and diffusion in suspensions of yukawa-type colloidal particles},\
  }\href@noop {} {\bibfield  {journal} {\bibinfo  {journal} {J. Chem. Phys.}\
  }\textbf {\bibinfo {volume} {135}},\ \bibinfo {pages} {154504} (\bibinfo
  {year} {2011})}\BibitemShut {NoStop}%
\bibitem [{\citenamefont {Westermeier}\ \emph {et~al.}(2012)\citenamefont
  {Westermeier}, \citenamefont {Fischer}, \citenamefont {Roseker},
  \citenamefont {Gr{\"u}bel}, \citenamefont {N{\"a}gele},\ and\ \citenamefont
  {Heinen}}]{westermeier2012structure}%
  \BibitemOpen
  \bibfield  {author} {\bibinfo {author} {\bibfnamefont {F.}~\bibnamefont
  {Westermeier}}, \bibinfo {author} {\bibfnamefont {B.}~\bibnamefont
  {Fischer}}, \bibinfo {author} {\bibfnamefont {W.}~\bibnamefont {Roseker}},
  \bibinfo {author} {\bibfnamefont {G.}~\bibnamefont {Gr{\"u}bel}}, \bibinfo
  {author} {\bibfnamefont {G.}~\bibnamefont {N{\"a}gele}},\ and\ \bibinfo
  {author} {\bibfnamefont {M.}~\bibnamefont {Heinen}},\ }\bibfield  {title}
  {\bibinfo {title} {Structure and short-time dynamics in concentrated
  suspensions of charged colloids},\ }\href@noop {} {\bibfield  {journal}
  {\bibinfo  {journal} {J. Chem. Phys.}\ }\textbf {\bibinfo {volume} {137}},\
  \bibinfo {pages} {114504} (\bibinfo {year} {2012})}\BibitemShut {NoStop}%
\bibitem [{\citenamefont {Riest}\ and\ \citenamefont
  {N{\"a}gele}(2015)}]{riest2015short}%
  \BibitemOpen
  \bibfield  {author} {\bibinfo {author} {\bibfnamefont {J.}~\bibnamefont
  {Riest}}\ and\ \bibinfo {author} {\bibfnamefont {G.}~\bibnamefont
  {N{\"a}gele}},\ }\bibfield  {title} {\bibinfo {title} {Short-time dynamics in
  dispersions with competing short-range attraction and long-range repulsion},\
  }\href@noop {} {\bibfield  {journal} {\bibinfo  {journal} {Soft Matter}\
  }\textbf {\bibinfo {volume} {11}},\ \bibinfo {pages} {9273} (\bibinfo {year}
  {2015})}\BibitemShut {NoStop}%
\bibitem [{\citenamefont {Rogers}\ \emph {et~al.}(2011)\citenamefont {Rogers},
  \citenamefont {Pagaduan}, \citenamefont {Nordin},\ and\ \citenamefont
  {Woolley}}]{Rogers2011}%
  \BibitemOpen
  \bibfield  {author} {\bibinfo {author} {\bibfnamefont {C.~I.}\ \bibnamefont
  {Rogers}}, \bibinfo {author} {\bibfnamefont {J.~V.}\ \bibnamefont
  {Pagaduan}}, \bibinfo {author} {\bibfnamefont {G.~P.}\ \bibnamefont
  {Nordin}},\ and\ \bibinfo {author} {\bibfnamefont {A.~T.}\ \bibnamefont
  {Woolley}},\ }\bibfield  {title} {\bibinfo {title} {Single-monomer
  formulation of polymerized polyethylene glycol diacrylate as a nonadsorptive
  material for microfluidics},\ }\href@noop {} {\bibfield  {journal} {\bibinfo
  {journal} {Anal. Chem.}\ }\textbf {\bibinfo {volume} {83}},\ \bibinfo {pages}
  {6418} (\bibinfo {year} {2011})}\BibitemShut {NoStop}%
\bibitem [{\citenamefont {Decock}\ \emph {et~al.}(2018)\citenamefont {Decock},
  \citenamefont {Schlenk},\ and\ \citenamefont {Salmon}}]{Decock2018}%
  \BibitemOpen
  \bibfield  {author} {\bibinfo {author} {\bibfnamefont {J.}~\bibnamefont
  {Decock}}, \bibinfo {author} {\bibfnamefont {M.}~\bibnamefont {Schlenk}},\
  and\ \bibinfo {author} {\bibfnamefont {J.-B.}\ \bibnamefont {Salmon}},\
  }\bibfield  {title} {\bibinfo {title} {In situ photo-patterning of
  pressure-resistant hydrogel membranes with controlled permeabilities in
  {PEGDA} microfluidic channels},\ }\href@noop {} {\bibfield  {journal}
  {\bibinfo  {journal} {Lab Chip}\ }\textbf {\bibinfo {volume} {18}},\ \bibinfo
  {pages} {1075} (\bibinfo {year} {2018})}\BibitemShut {NoStop}%
\bibitem [{\citenamefont {Lee}\ \emph {et~al.}(2010)\citenamefont {Lee},
  \citenamefont {Arena}, \citenamefont {Beebe},\ and\ \citenamefont
  {Palecek}}]{Lee:2010}%
  \BibitemOpen
  \bibfield  {author} {\bibinfo {author} {\bibfnamefont {A.~G.}\ \bibnamefont
  {Lee}}, \bibinfo {author} {\bibfnamefont {C.~P.}\ \bibnamefont {Arena}},
  \bibinfo {author} {\bibfnamefont {D.~J.}\ \bibnamefont {Beebe}},\ and\
  \bibinfo {author} {\bibfnamefont {S.~P.}\ \bibnamefont {Palecek}},\
  }\bibfield  {title} {\bibinfo {title} {Development of macroporous
  poly(ethylene glycol) hydrogel arrays within microfluidic channels},\
  }\href@noop {} {\bibfield  {journal} {\bibinfo  {journal}
  {Biomacromolecules}\ }\textbf {\bibinfo {volume} {11}},\ \bibinfo {pages}
  {3316} (\bibinfo {year} {2010})}\BibitemShut {NoStop}%
\bibitem [{\citenamefont {Shin}\ \emph {et~al.}(2016)\citenamefont {Shin},
  \citenamefont {Um}, \citenamefont {Sabass}, \citenamefont {Ault},
  \citenamefont {Rahimi}, \citenamefont {Warren},\ and\ \citenamefont
  {Stone}}]{Shin2016}%
  \BibitemOpen
  \bibfield  {author} {\bibinfo {author} {\bibfnamefont {S.}~\bibnamefont
  {Shin}}, \bibinfo {author} {\bibfnamefont {E.}~\bibnamefont {Um}}, \bibinfo
  {author} {\bibfnamefont {B.}~\bibnamefont {Sabass}}, \bibinfo {author}
  {\bibfnamefont {J.~T.}\ \bibnamefont {Ault}}, \bibinfo {author}
  {\bibfnamefont {M.}~\bibnamefont {Rahimi}}, \bibinfo {author} {\bibfnamefont
  {P.~B.}\ \bibnamefont {Warren}},\ and\ \bibinfo {author} {\bibfnamefont
  {H.~A.}\ \bibnamefont {Stone}},\ }\bibfield  {title} {\bibinfo {title}
  {Size-dependent control of colloid transport via solute gradients in dead-end
  channels},\ }\href@noop {} {\bibfield  {journal} {\bibinfo  {journal} {Proc.
  Natl. Acad. Sci. USA}\ }\textbf {\bibinfo {volume} {113}},\ \bibinfo {pages}
  {257} (\bibinfo {year} {2016})}\BibitemShut {NoStop}%
\bibitem [{\citenamefont {Hallez}\ and\ \citenamefont
  {Meireles}(2017)}]{Hallez2017}%
  \BibitemOpen
  \bibfield  {author} {\bibinfo {author} {\bibfnamefont {Y.}~\bibnamefont
  {Hallez}}\ and\ \bibinfo {author} {\bibfnamefont {M.}~\bibnamefont
  {Meireles}},\ }\bibfield  {title} {\bibinfo {title} {{Fast, Robust Evaluation
  of the Equation of State of Suspensions of Charge-Stabilized Colloidal
  Spheres}},\ }\href@noop {} {\bibfield  {journal} {\bibinfo  {journal}
  {Langmuir}\ }\textbf {\bibinfo {volume} {33}},\ \bibinfo {pages} {10051}
  (\bibinfo {year} {2017})}\BibitemShut {NoStop}%
\bibitem [{\citenamefont {Herman}\ and\ \citenamefont
  {Walz}(2015)}]{Herman2015}%
  \BibitemOpen
  \bibfield  {author} {\bibinfo {author} {\bibfnamefont {D.}~\bibnamefont
  {Herman}}\ and\ \bibinfo {author} {\bibfnamefont {J.~Y.}\ \bibnamefont
  {Walz}},\ }\bibfield  {title} {\bibinfo {title} {Adsorption and stabilizing
  effects of highly-charged latex nanoparticles in dispersions of
  weakly-charged silica colloids},\ }\href@noop {} {\bibfield  {journal}
  {\bibinfo  {journal} {J. Colloid Interface Sci.}\ }\textbf {\bibinfo {volume}
  {449}},\ \bibinfo {pages} {143} (\bibinfo {year} {2015})}\BibitemShut
  {NoStop}%
\bibitem [{\citenamefont {Alexander}\ \emph {et~al.}(1984)\citenamefont
  {Alexander}, \citenamefont {Chaikin}, \citenamefont {Grant}, \citenamefont
  {Morales}, \citenamefont {Pincus},\ and\ \citenamefont
  {Hone}}]{Alexander1984}%
  \BibitemOpen
  \bibfield  {author} {\bibinfo {author} {\bibfnamefont {S.}~\bibnamefont
  {Alexander}}, \bibinfo {author} {\bibfnamefont {P.~M.}\ \bibnamefont
  {Chaikin}}, \bibinfo {author} {\bibfnamefont {P.}~\bibnamefont {Grant}},
  \bibinfo {author} {\bibfnamefont {G.~J.}\ \bibnamefont {Morales}}, \bibinfo
  {author} {\bibfnamefont {P.}~\bibnamefont {Pincus}},\ and\ \bibinfo {author}
  {\bibfnamefont {D.}~\bibnamefont {Hone}},\ }\bibfield  {title} {\bibinfo
  {title} {{Charge renormalization, osmotic pressure, and bulk modulus of
  colloidal crystals: Theory}},\ }\href {https://doi.org/10.1063/1.446600}
  {\bibfield  {journal} {\bibinfo  {journal} {J. Chem. Phys.}\ }\textbf
  {\bibinfo {volume} {80}},\ \bibinfo {pages} {5776} (\bibinfo {year}
  {1984})}\BibitemShut {NoStop}%
\bibitem [{\citenamefont {Denton}(2010)}]{Denton2010}%
  \BibitemOpen
  \bibfield  {author} {\bibinfo {author} {\bibfnamefont {A.~R.}\ \bibnamefont
  {Denton}},\ }\bibfield  {title} {\bibinfo {title} {{Poisson-Boltzmann theory
  of charged colloids: limits of the cell model for salty suspensions}},\
  }\href@noop {} {\bibfield  {journal} {\bibinfo  {journal} {J. Phys.: Condens.
  Matter}\ }\textbf {\bibinfo {volume} {22}},\ \bibinfo {pages} {364108}
  (\bibinfo {year} {2010})}\BibitemShut {NoStop}%
\bibitem [{\citenamefont {Holm}\ \emph {et~al.}(2000)\citenamefont {Holm},
  \citenamefont {K{\'e}kicheff},\ and\ \citenamefont {Podgornik}}]{Holm2000}%
  \BibitemOpen
  \bibfield  {author} {\bibinfo {author} {\bibfnamefont {C.}~\bibnamefont
  {Holm}}, \bibinfo {author} {\bibfnamefont {P.}~\bibnamefont
  {K{\'e}kicheff}},\ and\ \bibinfo {author} {\bibfnamefont {R.}~\bibnamefont
  {Podgornik}},\ }\href@noop {} {\emph {\bibinfo {title} {{Electrostatic
  Effects in Soft Matter and Biophysics}}}},\ edited by\ \bibinfo {editor}
  {\bibfnamefont {C.}~\bibnamefont {Holm}}, \bibinfo {editor} {\bibfnamefont
  {P.}~\bibnamefont {K{\'e}kicheff}},\ and\ \bibinfo {editor} {\bibfnamefont
  {R.}~\bibnamefont {Podgornik}}\ (\bibinfo  {publisher} {NATO Science
  Series},\ \bibinfo {year} {2000})\BibitemShut {NoStop}%
\bibitem [{\citenamefont {Trizac}\ \emph {et~al.}(2007)\citenamefont {Trizac},
  \citenamefont {Belloni}, \citenamefont {Dobnikar}, \citenamefont {von
  Gr{\"u}nberg},\ and\ \citenamefont
  {Casta{\~n}eda-Priego}}]{trizac_macroion_2007}%
  \BibitemOpen
  \bibfield  {author} {\bibinfo {author} {\bibfnamefont {E.}~\bibnamefont
  {Trizac}}, \bibinfo {author} {\bibfnamefont {L.}~\bibnamefont {Belloni}},
  \bibinfo {author} {\bibfnamefont {J.}~\bibnamefont {Dobnikar}}, \bibinfo
  {author} {\bibfnamefont {H.~H.}\ \bibnamefont {von Gr{\"u}nberg}},\ and\
  \bibinfo {author} {\bibfnamefont {R.}~\bibnamefont {Casta{\~n}eda-Priego}},\
  }\bibfield  {title} {\bibinfo {title} {{Macroion virial contribution to the
  osmotic pressure in charge-stabilized colloidal suspensions}},\ }\href@noop
  {} {\bibfield  {journal} {\bibinfo  {journal} {Phys. Rev. E}\ }\textbf
  {\bibinfo {volume} {75}},\ \bibinfo {pages} {011401} (\bibinfo {year}
  {2007})}\BibitemShut {NoStop}%
\bibitem [{\citenamefont {Hallez}\ \emph {et~al.}(2014)\citenamefont {Hallez},
  \citenamefont {Diatta},\ and\ \citenamefont {Meireles}}]{Hallez2014}%
  \BibitemOpen
  \bibfield  {author} {\bibinfo {author} {\bibfnamefont {Y.}~\bibnamefont
  {Hallez}}, \bibinfo {author} {\bibfnamefont {J.}~\bibnamefont {Diatta}},\
  and\ \bibinfo {author} {\bibfnamefont {M.}~\bibnamefont {Meireles}},\
  }\bibfield  {title} {\bibinfo {title} {{Quantitative Assessment of the
  Accuracy of the Poisson--Boltzmann Cell Model for Salty Suspensions}},\
  }\href@noop {} {\bibfield  {journal} {\bibinfo  {journal} {Langmuir}\
  }\textbf {\bibinfo {volume} {30}},\ \bibinfo {pages} {6721} (\bibinfo {year}
  {2014})}\BibitemShut {NoStop}%
\bibitem [{\citenamefont {Fushiki}(1992)}]{fushiki1992molecular}%
  \BibitemOpen
  \bibfield  {author} {\bibinfo {author} {\bibfnamefont {M.}~\bibnamefont
  {Fushiki}},\ }\bibfield  {title} {\bibinfo {title} {Molecular-dynamics
  simulations for charged colloidal dispersions},\ }\href@noop {} {\bibfield
  {journal} {\bibinfo  {journal} {J. Chem. Phys.}\ }\textbf {\bibinfo {volume}
  {97}},\ \bibinfo {pages} {6700} (\bibinfo {year} {1992})}\BibitemShut
  {NoStop}%
\bibitem [{\citenamefont {Dobnikar}\ \emph {et~al.}(2004)\citenamefont
  {Dobnikar}, \citenamefont {Halo{\v{z}}an}, \citenamefont {Brumen},
  \citenamefont {Von~Gr{\"u}nberg},\ and\ \citenamefont
  {Rzehak}}]{dobnikar2004poisson}%
  \BibitemOpen
  \bibfield  {author} {\bibinfo {author} {\bibfnamefont {J.}~\bibnamefont
  {Dobnikar}}, \bibinfo {author} {\bibfnamefont {D.}~\bibnamefont
  {Halo{\v{z}}an}}, \bibinfo {author} {\bibfnamefont {M.}~\bibnamefont
  {Brumen}}, \bibinfo {author} {\bibfnamefont {H.-H.}\ \bibnamefont
  {Von~Gr{\"u}nberg}},\ and\ \bibinfo {author} {\bibfnamefont {R.}~\bibnamefont
  {Rzehak}},\ }\bibfield  {title} {\bibinfo {title} {Poisson-{Boltzmann
  Brownian} dynamics of charged colloids in suspension},\ }\href@noop {}
  {\bibfield  {journal} {\bibinfo  {journal} {Comput. Phys. Commun.}\ }\textbf
  {\bibinfo {volume} {159}},\ \bibinfo {pages} {73} (\bibinfo {year}
  {2004})}\BibitemShut {NoStop}%
\bibitem [{\citenamefont {Hallez}\ and\ \citenamefont
  {Meireles}(2016)}]{Hallez2016a}%
  \BibitemOpen
  \bibfield  {author} {\bibinfo {author} {\bibfnamefont {Y.}~\bibnamefont
  {Hallez}}\ and\ \bibinfo {author} {\bibfnamefont {M.}~\bibnamefont
  {Meireles}},\ }\bibfield  {title} {\bibinfo {title} {Modeling the
  electrostatics of hollow shell suspensions: Ion distribution, pair
  interactions, and many-body effects},\ }\href@noop {} {\bibfield  {journal}
  {\bibinfo  {journal} {Langmuir}\ }\textbf {\bibinfo {volume} {32}},\ \bibinfo
  {pages} {10430} (\bibinfo {year} {2016})}\BibitemShut {NoStop}%
\bibitem [{\citenamefont {Teulon}\ \emph {et~al.}(2019)\citenamefont {Teulon},
  \citenamefont {Hallez}, \citenamefont {Raffy}, \citenamefont {Guerin},
  \citenamefont {Palleau},\ and\ \citenamefont {Ressier}}]{Teulon2019}%
  \BibitemOpen
  \bibfield  {author} {\bibinfo {author} {\bibfnamefont {L.}~\bibnamefont
  {Teulon}}, \bibinfo {author} {\bibfnamefont {Y.}~\bibnamefont {Hallez}},
  \bibinfo {author} {\bibfnamefont {S.}~\bibnamefont {Raffy}}, \bibinfo
  {author} {\bibfnamefont {F.}~\bibnamefont {Guerin}}, \bibinfo {author}
  {\bibfnamefont {E.}~\bibnamefont {Palleau}},\ and\ \bibinfo {author}
  {\bibfnamefont {L.}~\bibnamefont {Ressier}},\ }\bibfield  {title} {\bibinfo
  {title} {Electrostatic directed assembly of colloidal microparticles assisted
  by convective flow},\ }\href@noop {} {\bibfield  {journal} {\bibinfo
  {journal} {J. Phys. Chem. C}\ }\textbf {\bibinfo {volume} {123}},\ \bibinfo
  {pages} {783} (\bibinfo {year} {2019})}\BibitemShut {NoStop}%
\bibitem [{\citenamefont {Wajnryb}\ \emph {et~al.}(2004)\citenamefont
  {Wajnryb}, \citenamefont {Szymczak},\ and\ \citenamefont
  {Cichocki}}]{wajnryb2004brownian}%
  \BibitemOpen
  \bibfield  {author} {\bibinfo {author} {\bibfnamefont {E.}~\bibnamefont
  {Wajnryb}}, \bibinfo {author} {\bibfnamefont {P.}~\bibnamefont {Szymczak}},\
  and\ \bibinfo {author} {\bibfnamefont {B.}~\bibnamefont {Cichocki}},\
  }\bibfield  {title} {\bibinfo {title} {Brownian dynamics: divergence of
  mobility tensor},\ }\href@noop {} {\bibfield  {journal} {\bibinfo  {journal}
  {Physica A}\ }\textbf {\bibinfo {volume} {335}},\ \bibinfo {pages} {339}
  (\bibinfo {year} {2004})}\BibitemShut {NoStop}%
\bibitem [{\citenamefont {Salmon}\ \emph {et~al.}(2017)\citenamefont {Salmon},
  \citenamefont {Doumenc},\ and\ \citenamefont
  {Guerrier}}]{Salmon2017humidity}%
  \BibitemOpen
  \bibfield  {author} {\bibinfo {author} {\bibfnamefont {J.-B.}\ \bibnamefont
  {Salmon}}, \bibinfo {author} {\bibfnamefont {F.}~\bibnamefont {Doumenc}},\
  and\ \bibinfo {author} {\bibfnamefont {B.}~\bibnamefont {Guerrier}},\
  }\bibfield  {title} {\bibinfo {title} {Humidity-insensitive water evaporation
  from molecular complex fluids},\ }\href@noop {} {\bibfield  {journal}
  {\bibinfo  {journal} {Phys. Rev. E}\ }\textbf {\bibinfo {volume} {96}},\
  \bibinfo {pages} {032612} (\bibinfo {year} {2017})}\BibitemShut {NoStop}%
\bibitem [{\citenamefont {Liu}\ \emph {et~al.}(2000)\citenamefont {Liu},
  \citenamefont {Fedkiw},\ and\ \citenamefont {Kang}}]{liu2000boundary}%
  \BibitemOpen
  \bibfield  {author} {\bibinfo {author} {\bibfnamefont {X.-D.}\ \bibnamefont
  {Liu}}, \bibinfo {author} {\bibfnamefont {R.~P.}\ \bibnamefont {Fedkiw}},\
  and\ \bibinfo {author} {\bibfnamefont {M.}~\bibnamefont {Kang}},\ }\bibfield
  {title} {\bibinfo {title} {A boundary condition capturing method for
  {Poisson's} equation on irregular domains},\ }\href@noop {} {\bibfield
  {journal} {\bibinfo  {journal} {J. Comput. Phys.}\ }\textbf {\bibinfo
  {volume} {160}},\ \bibinfo {pages} {151} (\bibinfo {year}
  {2000})}\BibitemShut {NoStop}%
\end{thebibliography}
%

\end{document}